\documentclass[fleqn,usenatbib,useAMS]{ajas_new}
\usepackage{multirow}
\usepackage{array}
\usepackage{physics}
\usepackage{caption}
\usepackage{natbib}
\usepackage{longtable}
\usepackage{multirow}
\usepackage{graphicx}
\usepackage{lastpage}
\usepackage{eso-pic}   
\AddToShipoutPictureBG*{%
  \AtPageUpperLeft{%
        \hspace{\dimexpr\paperwidth-4cm\relax} 
        \raisebox{-2.8cm}{\includegraphics[height=1.5cm]{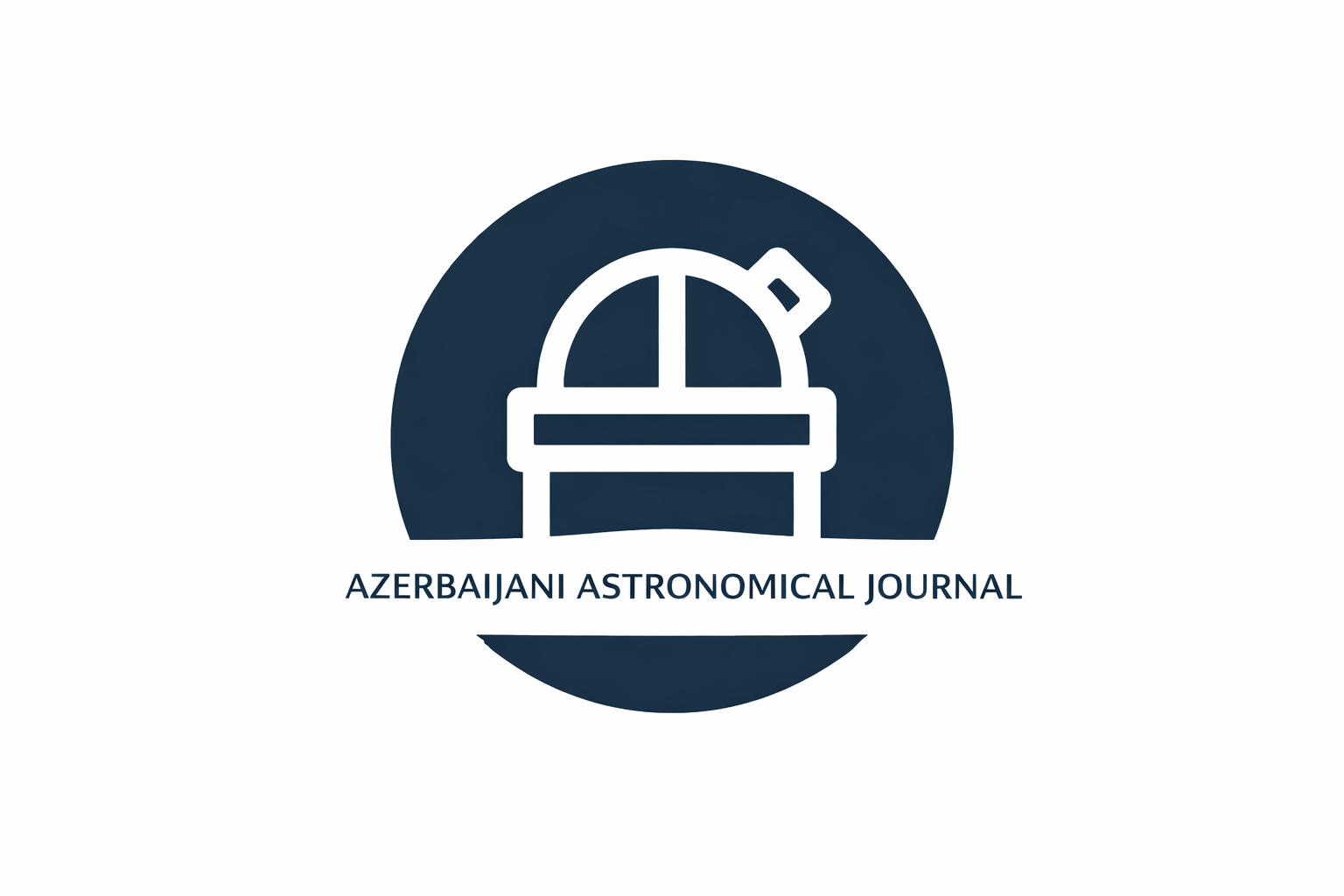}}%
      }%
    }%

\title[MAGNETIC CP~STARS] {REVERSIVE AND NON-REVERSIVE MAGNETIC CP~STARS. I. CATALOG.}
\author[I.~I. Romanyuk.]{
I.~I. Romanyuk.\thanks{E-mail: \href{mailto:roman@sao.ru}{roman@sao.ru}}
\\
\\
Special Astrophysical Observatory,  Russian Academy of Sciences, Nizhnii Arkhyz, 369167~Russia\\
}
\date{Last updated 2025 October 19; in original form 2025 November 28}

\pubyear{2026}

\begin{document}

\label{firstpage}
\pagerange{\pageref{firstpage}--\pageref{lastpage}}
\maketitle

\begin{abstract}
An analysis of the distribution of signs of the longitudinal magnetic field component $B_z$ is presented for a sample of 307~chemically peculiar (CP) stars with reliably measured magnetic fields. For each star, the root-mean-square field $B_{\rm rms}$ and the extreme values of $B_z$ are used to classify the field configuration as either reversive or non-reversive. Among the 173~non-reversive stars, 100~exhibit a predominantly negative $B_z$ sign, while only 73~display a positive sign, corresponding to an excess of negative polarities by a factor of about 1.37. A subsample of stars with at least five independent measurements (254~objects) yields a similar ratio of 1.38, confirming the robustness of the asymmetry. The observed predominance of negative $B_z$ values among non-reversive CP stars in the solar neighbourhood is statistically significant. Possible origins of this asymmetry remain unclear and will be addressed in future studies.
\end{abstract}


\section{Introduction}

\label{intro}
Magnetic fields are found in approximately 10\% of \mbox{OBA-type} stars on the main sequence. In terms of effective temperature, mass, and luminosity, these objects do not differ from normal stars, but exhibit significant anomalies in the chemical composition of their atmospheres: the abundances of some elements, in particular rare-earth elements, exceeds the solar value by up to three orders of magnitude (see, e.g., \citealp{Ghazaryan2018}). In addition, the rotation velocities of magnetic stars are approximately in 3--4 times lower than those of normal stars \citep{Netopil2017}.

The first magnetic field in a star (78\,Vir) was discovered by \citet{Babcock1947}. The technique he introduced allowed the measurement of only the longitudinal component $B_z$, under the assumption of a global, coherent field structure covering the entire stellar surface. To a first approximation, this configuration is dipolar. Although more complex geometries are possible, the dipole component remains dominant. The measured strengths of the large-scale magnetic fields in CP~stars range from about 100~G \citep{Auriere2007} up to 35~kG \citep{Babcock1960}.

The longitudinal component reaches its maximum value at the magnetic poles, and at the magnetic equator it becomes zero. As the star rotates, the observer views different regions of the surface with varying $B_z$ orientations, resulting in periodic changes of the measured field strength and polarity. This behaviour is naturally explained by the oblique rotator model \citep{Stibbs1950}. The choice of sign \mbox{($+$ or $-$)} is a matter of convention. Currently, researchers follow the system introduced by \citet{Babcock1947}. If during the rotation period the $B_z$ curve changes sign, the field is called reversive, and if the sign remains constant, then it is called non-reversive.

Since the rotation period is unknown for most of the stars studied and, therefore, it is impossible to carry out detailed modeling of the magnetic field, to estimate it we use the value of the root-mean-square magnetic field $B_{\rm rms}$ and its error:
\begin{equation}
	\label{Romanyuk_eq1}
	B_{\rm rms} =\sqrt{\frac{1}{n} \sum_{i=1}^n B_{{\rm z}i}^2}; \qquad 
    \sigma_{\rm rms} = \sqrt{\frac{1}{n} \sum_{i=1}^n \sigma_{i}^2}.
\end{equation}

\begin{center}
\small
\setlength\LTleft{\fill}
\setlength\LTright{\fill}
\setcounter{table}{0}
\begin{table*}
\renewcommand{\tabcolsep}{5.0pt}
\renewcommand{\baselinestretch}{0.75}
\caption{\space
Catalog of reversive and non-reversive magnetic CP stars. The stars are arranged in ascending order of their HD and BD catalog numbers. The table columns contain: star name, distance $r$, root-mean-square magnetic field and its error $B_{\rm rms} \pm \sigma$, minimum and maximum extremal values of the longitudinal field $B_z$, number of measurements $n$, and reversibility signature.}
\medskip
\label{Romanyuk_tab1}
\begin{tabular}{l|c|r@{$\,\pm\,$}l|r@{\,/\,}l|c|c||l|c|r@{$\,\pm\,$}l|r@{\,/\,}l|c|c}
\hline
\multicolumn{1}{c|}{\multirow{2}{*}{Star}} & $r$, & \multicolumn{2}{c|}{$B_{\rm rms} \pm \sigma$,} & \multicolumn{2}{c|}{$B_z\,{\rm (min)/(max)}$,} & $n$ & Revers. & \multicolumn{1}{c|}{\multirow{2}{*}{Star}} & $r$, & \multicolumn{2}{c|}{$B_{\rm rms} \pm \sigma$,} & \multicolumn{2}{c|}{$B_z\,{\rm (min)/(max)}$,} & $n$ & Revers. \\
  &   pc & \multicolumn{2}{c|}{G} & \multicolumn{2}{c|}{G} &   & type &   &   pc & \multicolumn{2}{c|}{G} & \multicolumn{2}{c|}{G} &   & type \\
\hline
HD\,315      	&	 219 	&	 1520  	&	  680  	&	 $-$1600 	&	 $+$2000 	&	 5  	&	 $+$/$-$	&	HD\,34736    	&	 373 	&	 4000  	&	  500  	&	 $-$6000 	&	 $+$5000 	&	 100 	&	 $+$/$-$	\\
HD\,965      	&	 237 	&	 400   	&	  50   	&	 $-$400  	&	 $+$600  	&	 78 	&	 $+$/$-$	&	HD\,34889    	&	 353 	&	 642   	&	  134  	&	 $-$1000 	&	 $+$600  	&	 6  	&	 $+$/$-$	\\
HD\,1048     	&	 139 	&	 136   	&	  55   	&	 $-$70   	&	 $+$240  	&	 4  	&	 $+$/$-$	&	HD\,349321   	&	 460 	&	 2700  	&	  300  	&	 $-$4400 	&	 $+$1900 	&	 20 	&	 $-$	\\
HD\,2453     	&	 174 	&	 588   	&	  202  	&	 $-$1030 	&	 $-$520  	&	 28 	&	 $-$	&	HD\,35177    	&	 347 	&	 940   	&	  274  	&	 $-$400  	&	 $+$700  	&	 4  	&	 $+$/$-$	\\
HD\,2887     	&	 331 	&	 436   	&	  249  	&	 $-$700  	&	 $+$500  	&	 6  	&	 $+$/$-$	&	HD\,35298    	&	 358 	&	 2323  	&	  330  	&	 $-$2810 	&	 $+$2920 	&	 13 	&	 $+$/$-$	\\
HD\,2957     	&	 316 	&	 520   	&	  120  	&	 $-$930  	&	 $+$670  	&	 7  	&	 $+$/$-$	&	HD\,35456    	&	 346 	&	 440   	&	  80   	&	 $+$200  	&	 $+$500  	&	 6  	&	 $+$	\\
HD\,3980     	&	 68  	&	 1200  	&	  200  	&	 $-$1600 	&	 $+$2000 	&	 11 	&	 $+$/$-$	&	HD\,35502    	&	 363 	&	 1490  	&	  140  	&	 $-$2250 	&	 $-$100  	&	 25 	&	 $-$	\\
HD\,4478     	&	 575 	&	 990   	&	  850  	&	 $-$1000 	&	 $+$1300 	&	 3  	&	 $+$/$-$	&	HD\,36313    	&	 386 	&	 1020  	&	  450  	&	 $-$1500 	&	 $+$1100 	&	 12 	&	 $+$/$-$	\\
HD\,4778     	&	 104 	&	 1026  	&	  454  	&	 $-$1100 	&	 $+$1400 	&	 30 	&	 $+$/$-$	&	HD\,36429    	&	 338 	&	 425   	&	  170  	&	 $-$840  	&	 $+$160  	&	 5  	&	 $-$	\\
HD\,5441     	&	 385 	&	 409   	&	  40   	&	 $-$440  	&	 0     	&	 6  	&	 $-$	&	HD\,36485    	&	 382 	&	 3220  	&	  318  	&	 $-$3700 	&	 $-$1900 	&	 10 	&	 $-$	\\
HD\,5601     	&	 277 	&	 1190  	&	  100  	&	 $-$2000 	&	 $-$300  	&	 8  	&	 $-$	&	HD\,36526    	&	 448 	&	 2130  	&	  200  	&	 $-$980  	&	 $+$3480 	&	 9  	&	 $+$	\\
HD\,5737     	&	 216 	&	 324   	&	  142  	&	 $-$400  	&	 $+$500  	&	 23 	&	 $+$/$-$	&	HD\,36644    	&	 338 	&	 1992  	&	  125  	&	 $-$2500 	&	 $+$2300 	&	 9  	&	 $+$/$-$	\\
HD\,5797     	&	 565 	&	 669   	&	  46   	&	 $-$100  	&	 $+$1000 	&	 10 	&	 $+$	&	HD\,36668    	&	 393 	&	 1875  	&	  447  	&	 $-$2100 	&	 $+$2000 	&	 10 	&	 $+$/$-$	\\
HD\,6757\,A    	&	 298 	&	 2730  	&	  160  	&	 $+$2300 	&	 $+$2900 	&	 30 	&	 $+$	&	HD\,36916    	&	 260 	&	 500   	&	  125  	&	 $-$1000 	&	 $-$200  	&	 8  	&	 $-$	\\
HD\,8855     	&	 485 	&	 354   	&	  145  	&	 $-$600  	&	 $+$270  	&	 6  	&	 $+$/$-$	&	HD\,36955    	&	 448 	&	 843   	&	  219  	&	 $-$1300 	&	 0     	&	 6  	&	 $-$	\\
HD\,9147     	&	 787 	&	 400   	&	  150  	&	 $-$370  	&	 $+$600  	&	 5  	&	 $+$/$-$	&	HD\,36997    	&	 424 	&	 762   	&	  73   	&	 $-$1600 	&	 $+$1100 	&	 6  	&	 $+$/$-$	\\
HD\,9996     	&	 140 	&	 833   	&	  174  	&	 $-$1700 	&	 $+$600  	&	 20 	&	 $-$	&	HD\,37017    	&	 360 	&	 1490  	&	  338  	&	 $-$2600 	&	 $+$300  	&	 44 	&	 $-$	\\
HD\,10783    	&	 169 	&	 1269  	&	  260  	&	 $-$1200 	&	 $+$1800 	&	 17 	&	 $+$/$-$	&	HD\,37058    	&	 382 	&	 1090  	&	  412  	&	 $-$1200 	&	 $+$900  	&	 11 	&	 $+$/$-$	\\
HD\,11187    	&	 389 	&	 616   	&	  239  	&	 $-$500  	&	 $+$400  	&	 11 	&	 $+$/$-$	&	HD\,37140    	&	 420 	&	 450   	&	  210  	&	 $-$900  	&	 $+$500  	&	 8  	&	 $+$/$-$	\\
HD\,11503    	&	 51  	&	 545   	&	  344  	&	 $-$1100 	&	 $+$500  	&	 18 	&	 $-$	&	HD\,37333    	&	 335 	&	 869   	&	  246  	&	 $-$1290 	&	 $+$560  	&	 6  	&	 $-$	\\
HD\,11948    	&	 180 	&	 468   	&	  39   	&	 $-$750  	&	 $+$330  	&	 8  	&	 $-$	&	HD\,37479    	&	 435 	&	 1980  	&	  155  	&	 $-$1600 	&	 $+$3500 	&	 15 	&	 $+$	\\
HD\,12288    	&	 251 	&	 1643  	&	  150  	&	 $-$3100 	&	 $-$200  	&	 42 	&	 $-$	&	HD\,37633    	&	 417 	&	 520   	&	  122  	&	 $+$200  	&	 $+$800  	&	 6  	&	 $+$	\\
HD\,12447    	&	 50  	&	 365   	&	  266  	&	 $-$510  	&	 $+$570  	&	 22 	&	 $+$/$-$	&	HD\,37642    	&	 369 	&	 2100  	&	  180  	&	 $-$3000 	&	 $+$3000 	&	 12 	&	 $+$/$-$	\\
HD\,12767    	&	 106 	&	 242   	&	  94   	&	 $-$230  	&	 $+$290  	&	 8  	&	 $+$/$-$	&	HD\,37687    	&	 446 	&	 613   	&	  195  	&	 $+$500  	&	 $+$700  	&	 5  	&	 $+$	\\
HD\,13404    	&	 217 	&	 3680  	&	  380  	&	 $-$4800 	&	 $-$2100 	&	 20 	&	 $-$	&	HD\,37776    	&	 394 	&	 1260  	&	  385  	&	 $-$2000 	&	 $+$2000 	&	 150 	&	 $+$/$-$	\\
HD\,14437    	&	 234 	&	 1829  	&	  260  	&	 $-$2600 	&	 $-$800  	&	 26 	&	 $-$	&	HD\,37808    	&	 185 	&	 950   	&	  172  	&	 $+$300  	&	 $+$1200 	&	 5  	&	 $+$	\\
HD\,15089    	&	 41  	&	 303   	&	  47   	&	 $-$300  	&	 $+$400  	&	 16 	&	 $+$/$-$	&	HD\,38129    	&	 178 	&	 438   	&	  173  	&	 $-$600  	&	 $-$200  	&	 3  	&	 $-$	\\
HD\,15144    	&	 79  	&	 802   	&	  216  	&	 $-$1150 	&	 $-$530  	&	 20 	&	 $-$	&	HD\,38823    	&	 110 	&	 1510  	&	  110  	&	 $-$2500 	&	 $+$1500 	&	 12 	&	 $+$/$-$	\\
HD\,16145    	&	 249 	&	 725   	&	  66   	&	 $-$820  	&	 $+$500  	&	 3  	&	 $+$/$-$	&	HD\,39082    	&	 166 	&	 1290  	&	  330  	&	 $-$1100 	&	 $+$1600 	&	 6  	&	 $+$/$-$	\\
HD\,16605    	&	 621 	&	 1627  	&	  119  	&	 $-$2400 	&	 $-$800  	&	 7  	&	 $-$	&	HD\,39658    	&	 371 	&	 930   	&	  190  	&	 $-$970  	&	 $+$1350 	&	 7  	&	 $+$/$-$	\\
HD\,16705    	&	 487 	&	 524   	&	  155  	&	 $-$400  	&	 $+$720  	&	 4  	&	 $+$/$-$	&	HD\,40142    	&	 308 	&	 700   	&	  50   	&	 $-$780  	&	 $+$780  	&	 4  	&	 $+$/$-$	\\
HD\,17330    	&	 427 	&	 415   	&	  38   	&	 $-$500  	&	 $-$300  	&	 30 	&	 $-$	&	HD\,40146    	&	 474 	&	 595   	&	  105  	&	 $-$650  	&	 $+$650  	&	 3  	&	 $+$/$-$	\\
HD\,18078    	&	 388 	&	 800   	&	  100  	&	 $-$850  	&	 $+$1200 	&	 25 	&	 $+$/$-$	&	HD\,40312    	&	 51  	&	 340   	&	  60   	&	 $-$240  	&	 $+$360  	&	 20 	&	 $+$/$-$	\\
HD\,18296    	&	 107 	&	 440   	&	  216  	&	 $-$1000 	&	 $+$1350 	&	 8  	&	 $+$/$-$	&	HD\,40652    	&	 377 	&	 781   	&	  61   	&	 $-$1200 	&	 $-$270  	&	 6  	&	 $-$	\\
HD\,19712    	&	 166 	&	 2211  	&	  232  	&	 $-$3800 	&	 $+$2000 	&	 13 	&	 $+$/$-$	&	HD\,40711    	&	 487 	&	 500   	&	  180  	&	 $-$650  	&	 $+$320  	&	 6  	&	 $-$	\\
HD\,21699    	&	 178 	&	 973   	&	  247  	&	 $-$800  	&	 $+$1600 	&	 8  	&	 $+$/$-$	&	HD\,40759    	&	 411 	&	 1990  	&	  240  	&	 $+$600  	&	 $+$2400 	&	 30 	&	 $+$	\\
HD\,22316    	&	 137 	&	 1250  	&	  348  	&	 $-$2200 	&	 $+$600  	&	 19 	&	 $-$	&	HD\,41403    	&	 283 	&	 330   	&	  150  	&	 $-$500  	&	 $+$450  	&	 12 	&	 $+$/$-$	\\
HD\,22470    	&	 117 	&	 733   	&	  408  	&	 $-$1100 	&	 $+$1200 	&	 12 	&	 $+$/$-$	&	HD\,42605    	&	 283 	&	 270   	&	  47   	&	 $-$480  	&	 $+$100  	&	 9  	&	 $-$	\\
HD\,22920    	&	 191 	&	 307   	&	  159  	&	 $+$200  	&	 $+$400  	&	 4  	&	 $+$	&	HD\,42616    	&	 231 	&	 620   	&	  238  	&	 $-$840  	&	 $-$440  	&	 4  	&	 $-$	\\
HD\,24155    	&	 161 	&	 803   	&	  170  	&	 $-$440  	&	 $+$1660 	&	 8  	&	 $+$	&	HD\,43819    	&	 246 	&	 560   	&	  100  	&	 $-$20   	&	 $+$740  	&	 4  	&	 $+$	\\
HD\,24712    	&	 48  	&	 802   	&	  171  	&	 $+$300  	&	 $+$1300 	&	 30 	&	 $+$	&	HD\,45107    	&	 492 	&	 300   	&	  150  	&	 $-$140  	&	 $+$570  	&	 8  	&	 $+$	\\
HD\,25267    	&	 100 	&	 241   	&	  91   	&	 $-$345  	&	 $-$15   	&	 7  	&	 $-$	&	HD\,45530    	&	 273 	&	 590   	&	  270  	&	 $-$650  	&	 $+$750  	&	 5  	&	 $+$/$-$	\\
HD\,25823    	&	 124 	&	 668   	&	  470  	&	 $-$100  	&	 $+$1200 	&	 18 	&	 $+$	&	HD\,45583    	&	 314 	&	 2880  	&	  600  	&	 $-$2500 	&	 $+$3800 	&	 30 	&	 $+$/$-$	\\
HD\,25999    	&	 158 	&	 1155  	&	  233  	&	 $-$900  	&	 $+$1400 	&	 6  	&	 $+$/$-$	&	HD\,47103    	&	 360 	&	 3526  	&	  365  	&	 $-$4000 	&	 $-$2700 	&	 8  	&	 $-$	\\
HD\,27309    	&	 88  	&	 1755  	&	  602  	&	 $-$4000 	&	 $-$200  	&	 25 	&	 $-$	&	HD\,47152    	&	 121 	&	 856   	&	  207  	&	 $-$600  	&	 $+$1500 	&	 4  	&	 $+$	\\
HD\,27404    	&	 165 	&	 1326  	&	  174  	&	 $-$3100 	&	 $+$1400 	&	 15 	&	 $-$	&	HD\,47756    	&	 277 	&	 420   	&	  120  	&	 $-$650  	&	 $+$150  	&	 7  	&	 $-$	\\
HD\,279021   	&	 257 	&	 954   	&	  151  	&	 $-$960  	&	 $+$1300 	&	 5  	&	 $+$/$-$	&	HD\,49223    	&	 714 	&	 390   	&	  210  	&	 $-$120  	&	 $+$590  	&	 6  	&	 $+$	\\
HD\,29009    	&	 239 	&	 480   	&	  217  	&	 $-$600  	&	 $+$900  	&	 10 	&	 $+$/$-$	&	HD\,49333    	&	 216 	&	 618   	&	  300  	&	 $-$800  	&	 $+$800  	&	 8  	&	 $+$/$-$	\\
HD\,290665   	&	 392 	&	 2024  	&	  53   	&	 $-$2900 	&	 $+$4000 	&	 10 	&	 $+$/$-$	&	HD\,49713    	&	 212 	&	 2560  	&	  460  	&	 $-$2880 	&	 $+$2200 	&	 9  	&	 $+$/$-$	\\
HD\,294046   	&	 355 	&	 495   	&	  163  	&	 $-$900  	&	 $+$330  	&	 4  	&	 $-$	&	HD\,49976    	&	 102 	&	 1490  	&	  360  	&	 $-$2000 	&	 $+$2200 	&	 18 	&	 $+$/$-$	\\
HD\,29762    	&	 392 	&	 409   	&	  50   	&	 $+$150  	&	 $+$560  	&	 4  	&	 $+$	&	HD\,50169    	&	 507 	&	 1220  	&	  220  	&	 $-$1500 	&	 $+$2000 	&	 30 	&	 $+$/$-$	\\
HD\,29925    	&	 365 	&	 1064  	&	  170  	&	 $-$1400 	&	 $-$200  	&	 10 	&	 $-$	&	HD\,50403    	&	 305 	&	 663   	&	  82   	&	 $-$330  	&	 $+$1100 	&	 5  	&	 $+$	\\
HD\,30466    	&	 195 	&	 1464  	&	  293  	&	 $+$1000 	&	 $+$2200 	&	 14 	&	 $+$	&	HD\,50461    	&	 257 	&	 1500  	&	  700  	&	 $-$2800 	&	 $+$2200 	&	 9  	&	 $+$/$-$	\\
HD\,32145    	&	 287 	&	 2000  	&	  293  	&	 $-$2100 	&	 $+$2400 	&	 7  	&	 $+$/$-$	&	HD\,51418    	&	 178 	&	 401   	&	  200  	&	 $-$460  	&	 $+$750  	&	 15 	&	 $+$/$-$	\\
HD\,32633    	&	 217 	&	 2760  	&	  263  	&	 $-$4100 	&	 $+$1700 	&	 50 	&	 $-$	&	HD\,51684    	&	 222 	&	 1410  	&	  65   	&	 $-$1800 	&	 $-$1100 	&	 5  	&	 $-$	\\
HD\,335238   	&	 385 	&	 1738  	&	  247  	&	 $-$3040 	&	 $+$2260 	&	 12 	&	 $+$/$-$	&	HD\,52628    	&	 220 	&	 2000  	&	  80   	&	 $-$2050 	&	 $+$2100 	&	 10 	&	 $+$/$-$	\\
HD\,341037   	&	 333 	&	 1100  	&	  48   	&	 $-$1400 	&	 $-$840  	&	 6  	&	 $-$	&	HD\,53081    	&	 770 	&	 450   	&	  100  	&	 $-$600  	&	 $+$600  	&	 5  	&	 $+$/$-$	\\
HD\,34162    	&	 363 	&	 440   	&	  100  	&	 $-$750  	&	 $+$190  	&	 6  	&	 $-$	&	HD\,54118    	&	 96  	&	 1030  	&	  256  	&	 $-$1600 	&	 $+$1600 	&	 7  	&	 $+$/$-$	\\
HD\,343872   	&	 526 	&	 2930  	&	  320  	&	 $-$700  	&	 $+$4500 	&	 40 	&	 $+$	&	HD\,54824    	&	 282 	&	 589   	&	  153  	&	 $-$860  	&	 $+$480  	&	 8  	&	 $+$/$-$	\\
HD\,34452    	&	 142 	&	 743   	&	  434  	&	 $-$300  	&	 $+$1000 	&	 13 	&	 $+$	&	HD\,55522    	&	 278 	&	 505   	&	  100  	&	 $+$38   	&	 $+$873  	&	 3  	&	 $+$	\\
HD\,34719    	&	 160 	&	 930   	&	  220  	&	 $-$1300 	&	 $+$1300 	&	 9  	&	 $+$/$-$	&	HD\,55755    	&	 555 	&	 2670  	&	  150  	&	 $-$3280 	&	 $+$2180 	&	 5  	&	 $+$/$-$	\\
\hline
\end{tabular} 
\end{table*}
\end{center}
\begin{center}
\small
\setlength\LTleft{\fill}
\setlength\LTright{\fill}
\setcounter{table}{0}
\begin{table*}
\renewcommand{\tabcolsep}{5.0pt}
\renewcommand{\baselinestretch}{0.75}
\caption{\space (Contd.)}
\medskip
\label{Romanyuk_tab1_1}
\begin{tabular}{l|c|r@{$\,\pm\,$}l|r@{\,/\,}l|c|c||l|c|r@{$\,\pm\,$}l|r@{\,/\,}l|c|c}
\hline
\multicolumn{1}{c|}{\multirow{2}{*}{Star}} & $r$, & \multicolumn{2}{c|}{$B_{\rm rms} \pm \sigma$,} & \multicolumn{2}{c|}{$B_z\,{\rm (min)/(max)}$,} & $n$ & Revers. & \multicolumn{1}{c|}{\multirow{2}{*}{Star}} & $r$, & \multicolumn{2}{c|}{$B_{\rm rms} \pm \sigma$,} & \multicolumn{2}{c|}{$B_z\,{\rm (min)/(max)}$,} & $n$ & Revers. \\
  &   pc & \multicolumn{2}{c|}{G} & \multicolumn{2}{c|}{G} &   & type &   &   pc & \multicolumn{2}{c|}{G} & \multicolumn{2}{c|}{G} &   & type \\
\hline
HD\,58260    	&	 629 	&	 2290  	&	  300  	&	 $+$2000 	&	 $+$2600 	&	 10 	&	 $+$	&	HD\,130559   	&	 73  	&	 1375  	&	  100  	&	 $-$1300 	&	 $-$200  	&	 12 	&	 $-$	\\
HD\,59435    	&	 518 	&	 800   	&	  50   	&	 $-$1200 	&	 $+$900  	&	 35 	&	 $+$/$-$	&	HD\,133029   	&	 157 	&	 2420  	&	  319  	&	 $+$1300 	&	 $+$3300 	&	 16 	&	 $+$	\\
HD\,61045    	&	 488 	&	 680   	&	  532  	&	 $-$160  	&	 $+$470  	&	 5  	&	 $+$/$-$	&	HD\,133652   	&	 122 	&	 1110  	&	  200  	&	 $-$2100 	&	 $+$700  	&	 9  	&	 $-$	\\
HD\,61468    	&	 515 	&	 1900  	&	  51   	&	 $-$2500 	&	 $-$1000 	&	 5  	&	 $-$	&	HD\,134214   	&	 93  	&	 458   	&	  150  	&	 $-$900  	&	 $-$200  	&	 20 	&	 $-$	\\
HD\,62140    	&	 95  	&	 1336  	&	  306  	&	 $-$2200 	&	 $+$3200 	&	 30 	&	 $+$/$-$	&	HD\,134793   	&	 180 	&	 770   	&	  250  	&	 $-$800  	&	 $+$900  	&	 6  	&	 $+$/$-$	\\
HD\,63347    	&	 175 	&	 831   	&	  139  	&	 $-$1100 	&	 $+$1100 	&	 11 	&	 $+$/$-$	&	HD\,135679   	&	 265 	&	 1025  	&	  43   	&	 $+$200  	&	 $+$1300 	&	 6  	&	 $+$	\\
HD\,63401    	&	 185 	&	 400   	&	  80   	&	 $-$650  	&	 $+$340  	&	 4  	&	 $+$/$-$	&	HD\,137193   	&	 264 	&	 680   	&	  220  	&	 $+$230  	&	 $+$970  	&	 4  	&	 $+$	\\
HD\,64486    	&	 104 	&	 855   	&	  513  	&	 $-$1300 	&	 $+$600  	&	 6  	&	 $-$	&	HD\,137509   	&	 196 	&	 1020  	&	  416  	&	 $-$1200 	&	 $+$2200 	&	 8  	&	 $+$/$-$	\\
HD\,64740    	&	 245 	&	 565   	&	  114  	&	 $-$870  	&	 $+$530  	&	 15 	&	 $+$/$-$	&	HD\,137909   	&	 35  	&	 750   	&	  100  	&	 $-$900  	&	 $+$1000 	&	 50 	&	 $+$/$-$	\\
HD\,65339    	&	 97  	&	 3200  	&	  440  	&	 $-$5400 	&	 $+$4200 	&	 100 	&	 $+$/$-$	&	HD\,137949   	&	 79  	&	 1500  	&	  120  	&	 $+$980  	&	 $+$1920 	&	 20 	&	 $+$	\\
HD\,66350    	&	 472 	&	 480   	&	  100  	&	 $-$660  	&	 $+$520  	&	 5  	&	 $+$/$-$	&	HD\,138633   	&	 302 	&	 277   	&	  25   	&	 $-$300  	&	 $+$300  	&	 5  	&	 $+$/$-$	\\
HD\,70331    	&	 467 	&	 2800  	&	  184  	&	 $-$3000 	&	 $-$2000 	&	 14 	&	 $-$	&	HD\,138777   	&	 355 	&	 1870  	&	  57   	&	 $+$1700 	&	 $+$2200 	&	 9  	&	 $+$	\\
HD\,71866    	&	 130 	&	 1680  	&	  236  	&	 $-$1600 	&	 $+$2000 	&	 10 	&	 $+$/$-$	&	HD\,140160   	&	 68  	&	 860   	&	  712  	&	 $-$1840 	&	 $+$760  	&	 4  	&	 $+$/$-$	\\
HD\,72295    	&	 370 	&	 200   	&	  130  	&	 $-$360  	&	 $+$110  	&	 8  	&	 $-$	&	HD\,142070   	&	 180 	&	 400   	&	  60   	&	 $-$700  	&	 $+$600  	&	 20 	&	 $+$/$-$	\\
HD\,72968    	&	 104 	&	 480   	&	  288  	&	 $-$700  	&	 $+$500  	&	 20 	&	 $+$/$-$	&	HD\,142301   	&	 144 	&	 2100  	&	  420  	&	 $-$4100 	&	 $+$1600 	&	 4  	&	 $-$	\\
HD\,73340    	&	 149 	&	 1644  	&	  218  	&	 $-$2300 	&	 $-$900  	&	 5  	&	 $-$	&	HD\,142502   	&	 431 	&	 547   	&	  36   	&	 $-$500  	&	 $+$660  	&	 6  	&	 $+$/$-$	\\
HD\,74521    	&	 154 	&	 812   	&	  141  	&	 $-$200  	&	 $+$1400 	&	 10 	&	 $+$	&	HD\,142554   	&	 448 	&	 1310  	&	  290  	&	 $-$2250 	&	 $+$1740 	&	 5  	&	 $+$/$-$	\\
HD\,78316    	&	 188 	&	 208   	&	  205  	&	 $-$640  	&	 $+$460  	&	 25 	&	 $+$/$-$	&	HD\,142990   	&	 142 	&	 1304  	&	  255  	&	 $-$2500 	&	 $+$600  	&	 20 	&	 $-$	\\
HD\,79158    	&	 190 	&	 672   	&	  226  	&	 $-$1500 	&	 $+$1000 	&	 20 	&	 $+$/$-$	&	HD\,143473   	&	 162 	&	 4292  	&	  362  	&	 $+$4200 	&	 $+$5100 	&	 6  	&	 $+$	\\
HD\,80316    	&	 135 	&	 410   	&	  120  	&	 $-$590  	&	 $+$649  	&	 5  	&	 $+$/$-$	&	HD\,144334   	&	 136 	&	 783   	&	  257  	&	 $-$1400 	&	 $+$500  	&	 20 	&	 $-$	\\
HD\,81009    	&	 145 	&	 1430  	&	  236  	&	 $-$100  	&	 $+$2500 	&	 10 	&	 $+$	&	HD\,147010   	&	 130 	&	 4032  	&	  402  	&	 $-$5500 	&	 $-$2500 	&	 20 	&	 $-$	\\
HD\,83368    	&	 71  	&	 576   	&	  264  	&	 $-$800  	&	 $+$800  	&	 12 	&	 $+$/$-$	&	HD\,148199   	&	 167 	&	 900   	&	  247  	&	 $-$900  	&	 $+$1700 	&	 12 	&	 $+$/$-$	\\
HD\,86170    	&	 273 	&	 257   	&	  88   	&	 $-$340  	&	 $-$150  	&	 4  	&	 $-$	&	HD\,149764   	&	 175 	&	 700   	&	  65   	&	 $-$1213 	&	 $+$30   	&	 3  	&	 $-$	\\
HD\,86592    	&	 153 	&	 1908  	&	  115  	&	 $-$260  	&	 $+$2670 	&	 10 	&	 $+$	&	HD\,149822   	&	 146 	&	 450   	&	  180  	&	 $-$690  	&	 $+$510  	&	 7  	&	 $+$/$-$	\\
HD\,89069    	&	 314 	&	 434   	&	  29   	&	 $-$700  	&	 $+$600  	&	 13 	&	 $+$/$-$	&	HD\,149911   	&	 119 	&	 580   	&	  32   	&	 $-$700  	&	 $+$300  	&	 5  	&	 $-$	\\
HD\,90044    	&	 106 	&	 740   	&	  373  	&	 $-$800  	&	 $+$700  	&	 6  	&	 $+$/$-$	&	HD\,151199   	&	 103 	&	 280   	&	  100  	&	 $-$400  	&	 $+$150  	&	 5  	&	 $-$	\\
HD\,92499    	&	 269 	&	 1000  	&	  140  	&	 $-$1255 	&	 $-$964  	&	 3  	&	 $-$	&	HD\,151965   	&	 175 	&	 2602  	&	  282  	&	 $-$3700 	&	 $-$550  	&	 8  	&	 $-$	\\
HD\,92664    	&	 153 	&	 803   	&	  40   	&	 $-$1300 	&	 $-$100  	&	 18 	&	 $-$	&	HD\,152107   	&	 55  	&	 1487  	&	  250  	&	 $+$500  	&	 $+$2000 	&	 50 	&	 $+$	\\
HD\,93507    	&	 327 	&	 2164  	&	  278  	&	 $+$1600 	&	 $+$2600 	&	 12 	&	 $+$	&	HD\,153882   	&	 161 	&	 1750  	&	  462  	&	 $-$1800 	&	 $+$3100 	&	 15 	&	 $+$	\\
HD\,93700    	&	 335 	&	 836   	&	  50   	&	 $-$1170 	&	 $-$490  	&	 9  	&	 $-$	&	HD\,154708   	&	 148 	&	 6000  	&	  50   	&	 $+$5764 	&	 $+$7530 	&	 3  	&	 $+$	\\
HD\,94603    	&	 450 	&	 518   	&	  69   	&	 $-$145  	&	 $+$900  	&	 5  	&	 $+$	&	HD\,158450   	&	 159 	&	 1570  	&	  180  	&	 $-$4400 	&	 $+$800  	&	 10 	&	 $-$	\\
HD\,94660    	&	 166 	&	 2352  	&	  265  	&	 $-$3300 	&	 $-$1800 	&	 6  	&	 $-$	&	HD\,159545   	&	 408 	&	 613   	&	  125  	&	 $-$360  	&	 $+$1100 	&	 4  	&	 $+$	\\
HD\,96237    	&	 350 	&	 750   	&	  83   	&	 $-$1000 	&	 $-$700  	&	 4  	&	 $-$	&	HD\,161321   	&	 140 	&	 573   	&	  118  	&	 $-$600  	&	 $+$900  	&	 6  	&	 $+$/$-$	\\
HD\,96446    	&	 465 	&	 1105  	&	  248  	&	 $-$2100 	&	 $-$1100 	&	 11 	&	 $-$	&	HD\,164258   	&	 116 	&	 800   	&	  300  	&	 $-$500  	&	 $+$1200 	&	 6  	&	 $+$/$-$	\\
HD\,98088    	&	 127 	&	 802   	&	  284  	&	 $-$1200 	&	 $+$1000 	&	 10 	&	 $+$/$-$	&	HD\,164827   	&	 324 	&	 1567  	&	  615  	&	 $-$2300 	&	 $+$1600 	&	 5  	&	 $+$/$-$	\\
HD\,99563    	&	 255 	&	 600   	&	  80   	&	 $-$680  	&	 $+$670  	&	 6  	&	 $+$/$-$	&	HD\,165474   	&	 171 	&	 470   	&	  100  	&	 $-$300  	&	 $+$600  	&	 20 	&	 $+$/$-$	\\
HD\,101065   	&	 109 	&	 2241  	&	  450  	&	 $-$2300 	&	 $-$1040 	&	 3  	&	 $-$	&	HD\,166473   	&	 140 	&	 2150  	&	  220  	&	 $-$2200 	&	 $-$600  	&	 3  	&	 $-$	\\
HD\,103192   	&	 95  	&	 204   	&	  104  	&	 $-$250  	&	 $-$100  	&	 4  	&	 $-$	&	HD\,168733   	&	 198 	&	 815   	&	  276  	&	 $-$800  	&	 $-$400  	&	 4  	&	 $-$	\\
HD\,107000   	&	 277 	&	 200   	&	  60   	&	 $-$240  	&	 $+$390  	&	 15 	&	 $+$/$-$	&	HD\,168796   	&	 215 	&	 610   	&	  100  	&	 $-$900  	&	 $+$500  	&	 7  	&	 $+$/$-$	\\
HD\,107612   	&	 176 	&	 320   	&	  140  	&	 $-$440  	&	 $+$430  	&	 7  	&	 $+$/$-$	&	HD\,168856   	&	 204 	&	 657   	&	  221  	&	 $-$1100 	&	 $+$360  	&	 4  	&	 $-$	\\
HD\,108651   	&	 72  	&	 380   	&	  240  	&	 $-$200  	&	 $+$560  	&	 5  	&	 $+$	&	HD\,169842   	&	 403 	&	 370   	&	  180  	&	 $-$660  	&	 $+$380  	&	 8  	&	 $+$/$-$	\\
HD\,108662   	&	 77  	&	 620   	&	  200  	&	 $-$1150 	&	 $+$550  	&	 50 	&	 $-$	&	HD\,169887   	&	 376 	&	 680   	&	  250  	&	 $-$2300 	&	 $+$2020 	&	 9  	&	 $+$/$-$	\\
HD\,108945   	&	 82  	&	 537   	&	  313  	&	 $-$347  	&	 $+$440  	&	 15 	&	 $+$/$-$	&	HD\,170000   	&	 93  	&	 350   	&	  150  	&	 $-$180  	&	 $+$640  	&	 15 	&	 $+$	\\
HD\,109026   	&	 116 	&	 342   	&	  95   	&	 $+$140  	&	 $+$470  	&	 5  	&	 $+$	&	HD\,170397   	&	 113 	&	 615   	&	  252  	&	 $-$650  	&	 $+$870  	&	 10 	&	 $+$/$-$	\\
HD\,110066   	&	 137 	&	 204   	&	  50   	&	 $-$370  	&	 $+$300  	&	 15 	&	 $+$/$-$	&	HD\,170565   	&	 235 	&	 1760  	&	  170  	&	 $+$40   	&	 $+$1960 	&	 6  	&	 $+$	\\
HD\,111133   	&	 192 	&	 806   	&	  143  	&	 $-$1500 	&	 $-$500  	&	 10 	&	 $-$	&	HD\,170836   	&	 666 	&	 490   	&	  140  	&	 $-$700  	&	 $+$300  	&	 3  	&	 $-$	\\
HD\,112381   	&	 124 	&	 3400  	&	  245  	&	 $-$3700 	&	 $-$3100 	&	 5  	&	 $-$	&	HD\,170973   	&	 251 	&	 530   	&	  100  	&	 $-$600  	&	 $+$800  	&	 8  	&	 $+$/$-$	\\
HD\,112413   	&	 30  	&	 1350  	&	  200  	&	 $-$1400 	&	 $+$1600 	&	 100 	&	 $+$/$-$	&	HD\,173650   	&	 247 	&	 326   	&	  275  	&	 $-$500  	&	 $+$700  	&	 15 	&	 $+$/$-$	\\
HD\,112528   	&	 226 	&	 900   	&	  100  	&	 $-$1000 	&	 $+$970  	&	 8  	&	 $+$/$-$	&	HD\,175362   	&	 146 	&	 3570  	&	  448  	&	 $-$6500 	&	 $+$4000 	&	 20 	&	 $+$/$-$	\\
HD\,113894   	&	 192 	&	 791   	&	  40   	&	 $-$1200 	&	 $+$1000 	&	 11 	&	 $+$/$-$	&	HD\,178892   	&	 249 	&	 5410  	&	  470  	&	 $+$2100 	&	 $+$8000 	&	 32 	&	 $+$	\\
HD\,115605   	&	 181 	&	 620   	&	  120  	&	 $-$750  	&	 $+$680  	&	 7  	&	 $+$/$-$	&	HD\,179761   	&	 210 	&	 480   	&	  238  	&	 $-$590  	&	 $+$170  	&	 4  	&	 $-$	\\
HD\,115708   	&	 128 	&	 927   	&	  405  	&	 $-$1860 	&	 $+$1600 	&	 11 	&	 $+$/$-$	&	HD\,180058   	&	 641 	&	 277   	&	  74   	&	 $-$390  	&	 $+$310  	&	 4  	&	 $+$/$-$	\\
HD\,116114   	&	 138 	&	 1923  	&	  113  	&	 $-$2200 	&	 $-$1800 	&	 13 	&	 $-$	&	HD\,182532   	&	 316 	&	 461   	&	  108  	&	 $-$40   	&	 $+$620  	&	 5  	&	 $+$	\\
HD\,116458   	&	 125 	&	 1925  	&	  273  	&	 $-$2300 	&	 $-$1500 	&	 10 	&	 $-$	&	HD\,183339   	&	 322 	&	 1296  	&	  465  	&	 $-$1600 	&	 $+$1800 	&	 6  	&	 $+$/$-$	\\
HD\,118022   	&	 60  	&	 808   	&	  225  	&	 $-$2800 	&	 $-$200  	&	 80 	&	 $-$	&	HD\,184471   	&	 460 	&	 350   	&	  100  	&	 $-$20   	&	 $+$800  	&	 15 	&	 $+$	\\
HD\,118054   	&	 148 	&	 575   	&	  304  	&	 $-$1000 	&	 $+$200  	&	 6  	&	 $-$	&	HD\,184927   	&	 599 	&	 1465  	&	  430  	&	 $-$1200 	&	 $+$3000 	&	 15 	&	 $+$	\\
HD\,119213   	&	 91  	&	 1220  	&	  440  	&	 $-$500  	&	 $+$1600 	&	 13 	&	 $+$	&	HD\,187128   	&	 336 	&	 400   	&	  66   	&	 $+$373  	&	 $+$437  	&	 3  	&	 $+$	\\
HD\,119419   	&	 134 	&	 1770  	&	  455  	&	 $-$4200 	&	 $+$1800 	&	 20 	&	 $-$	&	HD\,187474   	&	 97  	&	 1488  	&	  143  	&	 $-$1800 	&	 $+$1800 	&	 10 	&	 $+$/$-$	\\
HD\,122532   	&	 153 	&	 665   	&	  268  	&	 $-$900  	&	 $+$900  	&	 9  	&	 $+$/$-$	&	HD\,188041   	&	 85  	&	 1100  	&	  200  	&	 $-$200  	&	 $+$1500 	&	 10 	&	 $+$	\\
HD\,124224   	&	 75  	&	 570   	&	  323  	&	 $-$437  	&	 $+$811  	&	 14 	&	 $+$/$-$	&	HD\,188501   	&	 462 	&	 1172  	&	  72   	&	 $-$2200 	&	 $+$50   	&	 11 	&	 $-$	\\
HD\,125248   	&	 87  	&	 1505  	&	  295  	&	 $-$2500 	&	 $+$2800 	&	 15 	&	 $+$/$-$	&	HD\,189160   	&	 332 	&	 995   	&	  197  	&	 $-$200  	&	 $+$1500 	&	 4  	&	 $+$	\\
HD\,125823   	&	 120 	&	 470   	&	  253  	&	 $-$440  	&	 $+$370  	&	 9  	&	 $+$/$-$	&	HD\,189775   	&	 248 	&	 1375  	&	  540  	&	 $-$540  	&	 $+$2240 	&	 3  	&	 $+$	\\
HD\,126515   	&	 131 	&	 1720  	&	  373  	&	 $-$2000 	&	 $+$2000 	&	 15 	&	 $+$/$-$	&	HD\,189963   	&	 621 	&	 410   	&	  130  	&	 $-$700  	&	 $+$300  	&	 6  	&	 $-$	\\
\hline
\end{tabular} 
\end{table*}
\end{center}
\small
\setlength\LTleft{\fill}
\setlength\LTright{\fill}
\setcounter{table}{0}
\begin{table*}
\renewcommand{\tabcolsep}{4.0pt}
\renewcommand{\baselinestretch}{0.75}
\caption{(Contd.)}
\medskip
\label{Romanyuk_tab1_2}
\begin{tabular}{l|c|r@{$\,\pm\,$}l|r@{\,/\,}l|c|c||l|c|r@{$\,\pm\,$}l|r@{\,/\,}l|c|c}
\hline
\multicolumn{1}{c|}{\multirow{2}{*}{Star}} & $r$, & \multicolumn{2}{c|}{$B_{\rm rms} \pm \sigma$,} & \multicolumn{2}{c|}{$B_z\,{\rm (min)/(max)}$,} & $n$ & Revers. & \multicolumn{1}{c|}{\multirow{2}{*}{Star}} & $r$, & \multicolumn{2}{c|}{$B_{\rm rms} \pm \sigma$,} & \multicolumn{2}{c|}{$B_z\,{\rm (min)/(max)}$,} & $n$ & Revers. \\
  &   pc & \multicolumn{2}{c|}{G} & \multicolumn{2}{c|}{G} &   & type &   &   pc & \multicolumn{2}{c|}{G} & \multicolumn{2}{c|}{G} &   & type \\
\hline
HD\,191742   	&	 329 	&	 578   	&	  43   	&	 $-$900  	&	 $-$200  	&	 4  	&	 $-$	&	HD\,225114   	&	 277 	&	 970   	&	  220  	&	 $-$1450 	&	 $+$750  	&	 20 	&	 $+$/$-$	\\
HD\,192224   	&	 440 	&	 480   	&	  130  	&	 $-$680  	&	 $+$680  	&	 8  	&	 $+$/$-$	&	HD\,225627   	&	 393 	&	 204   	&	  59   	&	 $-$50   	&	 $+$300  	&	 5  	&	 $+$	\\
HD\,192678   	&	 220 	&	 1410  	&	  160  	&	 $+$1000 	&	 $+$1800 	&	 30 	&	 $+$	&	HD\,231054   	&	 321 	&	 1650  	&	  240  	&	 $+$380  	&	 $+$2500 	&	 5  	&	 $+$	\\
HD\,192913   	&	 242 	&	 483   	&	  221  	&	 $-$670  	&	 $+$380  	&	 5  	&	 $+$/$-$	&	HD\,258686   	&	 629 	&	 6000  	&	  320  	&	 $+$5100 	&	 $+$7900 	&	 17 	&	 $+$	\\
HD\,196178   	&	 168 	&	 973   	&	  238  	&	 $-$1500 	&	 $-$700  	&	 9  	&	 $-$	&	HD\,260858   	&	 790 	&	 536   	&	  76   	&	 $+$300  	&	 $+$700  	&	 4  	&	 $+$	\\
HD\,196502   	&	 116 	&	 490   	&	  200  	&	 $-$700  	&	 $-$200  	&	 15 	&	 $-$	&	HD\,279021   	&	 257 	&	 954   	&	  151  	&	 $-$960  	&	 $+$1300 	&	 5  	&	 $+$/$-$	\\
HD\,196606   	&	 251 	&	 1005  	&	  192  	&	 $-$1565 	&	 $+$1040 	&	 5  	&	 $+$/$-$	&	HD\,290665   	&	 392 	&	 2024  	&	  53   	&	 $-$2900 	&	 $+$4000 	&	 10 	&	 $+$/$-$	\\
HD\,196655   	&	 323 	&	 400   	&	  100  	&	 $-$530  	&	 $+$450  	&	 4  	&	 $+$/$-$	&	HD\,294046   	&	 355 	&	 495   	&	  163  	&	 $-$900  	&	 $+$330  	&	 4  	&	 $-$	\\
HD\,196691   	&	 371 	&	 1650  	&	  250  	&	 $-$800  	&	 $+$2290 	&	 5  	&	 $+$	&	HD\,335238   	&	 385 	&	 1738  	&	  247  	&	 $-$3040 	&	 $+$2260 	&	 12 	&	 $+$/$-$	\\
HD\,199180   	&	 385 	&	 278   	&	  45   	&	 $-$350  	&	 $-$5    	&	 7  	&	 $-$	&	HD\,338226   	&	 550 	&	 1080  	&	  200  	&	 $+$440  	&	 $+$1490 	&	 7  	&	 $+$	\\
HD\,199728   	&	 213 	&	 400   	&	  200  	&	 $-$470  	&	 $+$1120 	&	 6  	&	 $+$	&	HD\,341037   	&	 333 	&	 1100  	&	  48   	&	 $-$1400 	&	 $-$840  	&	 6  	&	 $-$	\\
HD\,200177   	&	 156 	&	 1124  	&	  433  	&	 $-$1900 	&	 $+$300  	&	 4  	&	 $-$	&	HD\,343872   	&	 526 	&	 2930  	&	  320  	&	 $-$700  	&	 $+$4500 	&	 40 	&	 $+$	\\
HD\,200311   	&	 330 	&	 1490  	&	  427  	&	 $-$2600 	&	 $+$2600 	&	 22 	&	 $+$/$-$	&	HD\,349321   	&	 460 	&	 2700  	&	  300  	&	 $-$4400 	&	 $+$1900 	&	 20 	&	 $-$	\\
HD\,200405   	&	 370 	&	 194   	&	  24   	&	 $-$265  	&	 $-$50   	&	 4  	&	 $-$	&	HDE\,293764   	&	 385 	&	 3760  	&	  220  	&	 $+$2700 	&	 $+$4700 	&	 16 	&	 $+$	\\
HD\,201174   	&	 305 	&	 1311  	&	  75   	&	 $+$500  	&	 $+$2100 	&	 40 	&	 $+$	&	BD\,$+$00\,1659 	&	 862 	&	 288   	&	  99   	&	 $+$160  	&	 $+$365  	&	 6  	&	 $+$	\\
HD\,201601   	&	 35  	&	 600   	&	  100  	&	 $-$1100 	&	 $+$600  	&	 100 	&	 $+$/$-$	&	BD\,$+$00\,4535 	&	 526 	&	 2157  	&	  180  	&	 $-$2900 	&	 $+$2400 	&	 6  	&	 $+$/$-$	\\
HD\,204815   	&	 512 	&	 390   	&	  145  	&	 $+$60   	&	 $+$600  	&	 5  	&	 $+$	&	BD\,$+$17\,3622 	&	 315 	&	 1390  	&	  180  	&	 $+$900  	&	 $+$1600 	&	 4  	&	 $+$	\\
HD\,205198   	&	 304 	&	 1714  	&	  557  	&	 $-$3700 	&	 $-$222  	&	 5  	&	 $-$	&	BD\,$+$32\,2827 	&	 568 	&	 524   	&	  156  	&	 $-$770  	&	 $+$60   	&	 3  	&	 $-$	\\
HD\,207188   	&	 293 	&	 1220  	&	  310  	&	 $-$1510 	&	 $+$1000 	&	 4  	&	 $+$/$-$	&	BD\,$+$35\,3616 	&	 417 	&	 520   	&	  160  	&	 $-$500  	&	 $+$540  	&	 6  	&	 $+$/$-$	\\
HD\,208217   	&	 122 	&	 970   	&	  45   	&	 $-$1800 	&	 $+$800  	&	 6  	&	 $-$	&	BD\,$+$40\,175A 	&	 300 	&	 2650  	&	  178  	&	 $-$3400 	&	 $-$2000 	&	 8  	&	 $-$	\\
HD\,209051   	&	 336 	&	 2225  	&	  520  	&	 $-$3300 	&	 $-$1040 	&	 7  	&	 $-$	&	BD\,$+$40\,175B 	&	 300 	&	 1426  	&	  148  	&	 $+$780  	&	 $+$2660 	&	 6  	&	 $+$	\\
HD\,210432   	&	 232 	&	 1230  	&	  190  	&	 $-$1810 	&	 $+$530  	&	 4  	&	 $-$	&	BD\,$+$41\,43   	&	 850 	&	 290   	&	  150  	&	 $-$450  	&	 $+$100  	&	 11 	&	 $-$	\\
HD\,213258   	&	 111 	&	 950   	&	  49   	&	 $-$1200 	&	 $-$400  	&	 26 	&	 $-$	&	BD\,$+$42\,659  	&	 427 	&	 913   	&	  68   	&	 $-$1000 	&	 $+$1300 	&	 4  	&	 $+$/$-$	\\
HD\,215441   	&	 460 	&	 17500 	&	  500  	&	 $+$10000 	&	 $+$20000 	&	 50 	&	 $+$	&	BD\,$+$44\,4130 	&	 540 	&	 2728  	&	  132  	&	 $-$3094 	&	 $-$2398 	&	 4  	&	 $-$	\\
HD\,216018   	&	 143 	&	 1200  	&	  120  	&	 $+$1000 	&	 $+$1400 	&	 12 	&	 $+$	&	BD\,$+$46\,570  	&	 474 	&	 450   	&	  68   	&	 $+$260  	&	 $+$570  	&	 4  	&	 $+$	\\
HD\,217833   	&	 204 	&	 3000  	&	  500  	&	 $-$6200 	&	 $-$1500 	&	 12 	&	 $-$	&	BD\,$+$51\,3356 	&	 507 	&	 3209  	&	  598  	&	 $-$218  	&	 $+$4786 	&	 3  	&	 $+$	\\
HD\,221006   	&	 119 	&	 600   	&	  150  	&	 $+$410  	&	 $+$990  	&	 3  	&	 $+$	&	BD\,$+$53\,1183 	&	 595 	&	 703   	&	  111  	&	 $-$810  	&	 $+$1030 	&	 7  	&	 $+$/$-$	\\
HD\,221934   	&	 312 	&	 1100  	&	  400  	&	 $-$1490 	&	 $-$1130 	&	 4  	&	 $-$	&	BD\,$+$64\,352  	&	 505 	&	 4240  	&	  390  	&	 $-$5500 	&	 $-$2700 	&	 17 	&	 $-$	\\
HD\,221936   	&	 444 	&	 2200  	&	  300  	&	 $-$2500 	&	 $+$2900 	&	 27 	&	 $+$/$-$	&	V1356\,Ori     	&	 917 	&	 2282  	&	  800  	&	 $-$3200 	&	 $-$800  	&	 6  	&	 $-$	\\
HD\,223640   	&	 100 	&	 653   	&	  218  	&	 $-$20   	&	 $+$820  	&	 4  	&	 $+$	&		&		&		&		&		&		&		&		\\
\hline
\end{tabular} 
\end{table*}
\renewcommand{\baselinestretch}{1.0}

The purpose of this study is to search for a possible correlation between the strength and configuration of the magnetic field of individual stars and the parameters of the large-scale Galactic magnetic field in the solar neighbourhood.

The first attempt at such an analysis was made by \citet{Romanyuk1994}. Based on 64~non-reversible CP stars, it was found that in 36~objects the negative sign of the longitudinal field predominates, and in 28~it is positive. However, due to the small sample size, it was premature to conclude that a genuine asymmetry exists in the distribution of magnetic stars. A substantial increase in the number of studied objects was clearly required.

Over the next 30~years, our group conducted an active search for new magnetic stars in the northern hemisphere. In total, about 200~objects were discovered. In parallel with these efforts, large international collaborations have carried out a number of survey projects, leading to the discovery of hundreds of new magnetic CP stars. Notable among these are the BinaMIcS (Binarity and Magnetic Interactions in various classes of Stars; \citealp{Neiner2015}), MiMeS (Magnetism in Massive Stars; \citealp{Wade2016}), and BOB (B-fields in OB stars; \citealp{Scholler2017}) projects. Thus, it became possible to perform analysis at a new level, using much more extensive observational material.

\section{CATALOG AND SAMPLE ANALYSIS}

Currently, magnetic fields have been detected in a significant number of CP stars. However, for many objects only a few measurements have been made, which does not allow constructing a phase magnetic curve and confidently classifying the field as reversible or non-reversible. For this reason, a list of stars was compiled for which there are at least three measurements of the longitudinal component of the magnetic field.

The primary data sources were the catalogs \citet{Romanyuk2008}, \citet{Bychkov2009}, together with the later compilation by \citet{Bychkov2021}. Additionally, the database of our group was used. The final catalog includes only those objects for which the root-mean-square value $B_{\rm rms}$ confirms the presence of a magnetic field with high confidence.

The resulting sample of reversive and non-reversive magnetic CP stars is presented in Table~\ref{Romanyuk_tab1}. In addition to the magnetic field parameters, it contains distances to objects that were obtained from the parallaxes of the Gaia~DR3 mission \citep{Gaia2020}. Almost all magnetic stars from catalog are nearby objects ($r \leq 1$~kpc). In a few cases where the Gaia parallaxes for binary systems are unreliable, distances determined from spectroscopic data. A star is classified as non-reversive if one of the extreme values of the longitudinal field exceeds the other by at least a factor of two or more times. For reversive fields in the table, ($+/-$) is indicated in the corresponding column.

In total, the catalogue contains 307~magnetic CP~stars. A reversive field is observed in 134~objects (44\%), while the remaining 173~stars (56\%) exhibit a non-reversive configuration. Among the latter, 73~objects have a positive field sign ($+$) and 100~have a negative sign ($-$). Thus, in the non-reversive subsample the ratio of negative to positive polarities is approximately 1.37. In the earlier study of \citet{Romanyuk1994}, based on a smaller sample, this ratio was about 1.28.

To classify a star as reversive or non-reversive, it is necessary to have a sufficient number of measurements of the longitudinal field. Obviously, three observations are not enough to confidently construct the $B_z$ curve and determine the alternation of signs. In order to check the stability of the obtained conclusions, stars were selected from the catalog for which at least five magnetic field measurements were made. For the majority of these objects, even in the absence of a known rotation period, five measurements provide a reliable characterisation of the field behaviour, and the probability of misclassification is greatly reduced.

A separate issue concerns super-slowly rotating stars, for which all measurements can occur at the same rotation phase. However, statistical estimates indicate that such objects constitute only a few percent of the total CP~star population (see, e.g., \citealp{Mathys2020}), and therefore they cannot significantly affect the overall conclusions.

After applying the selection criterion ($n\ge 5$), a total of 254~magnetic CP~stars remain. Among these, 77~are non-reversive stars with a negative sign ($-$), 56~have a positive sign ($+$), and 121~display reversive fields ($+/-$). The ratio of negative to positive polarities in this restricted subsample is 1.38, in excellent agreement with the value obtained from the full catalogue. Thus, in all the cases considered, the advantage of stars with a negative sign of the longitudinal field remains.

All stars are located within 1~kpc from the Sun, with the absolute majority (90\%) located at a distance of less than 500~pc. The nearest object is HD\,112413 at 30~pc, while the most distant is V1356\,Ori at 917~pc. The mean distance of the sample is $\langle r\rangle = 275$~pc.

The root-mean-square magnetic field $B_{\rm rms}$ varies in a wide range, from 200~G (HD\,72295 and HD\,107000) to 17\,500~G (HD\,215441), with a mean value of $\langle B_{\rm rms}\rangle = 1290 \pm 220$~G. The $B_{\rm rms}$ distribution shows that the field strengths of most stars do not exceed a few kilogauss: only 15~objects (6\%) have $B_{\rm rms} \ge 3$~kG.

\section{CONCLUSIONS}

This paper presents a catalogue of magnetic CP~stars containing 307~objects with reliably measured values of the longitudinal magnetic field $B_z$. All stars lie relatively close to the Sun, with distances not exceeding 1~kpc. For each object, the catalogue lists the root-mean-square magnetic field $B_{\rm rms}$, the extreme values of $B_z$, the number of observations, and the classification of the field as reversive or non-reversive.

An analysis of the full sample of 307~stars reveals that among non-reversive objects, those with a negative longitudinal field sign are approximately 37\% more common than those with a positive sign. This excess is larger than the value reported in the earlier study of \citet{Romanyuk1994}. A subsample restricted to 254~stars with five or more observations yields a similar predominance of the negative sign, confirming the robustness of the detected asymmetry.

On the basis of this statistically significant sample, it is concluded that non-reversive magnetic CP~stars in the solar neighbourhood possess a predominantly negative longitudinal magnetic field. The physical origin of this asymmetry remains unclear. Possible explanations include a connection with the orientation of the large-scale Galactic magnetic field in the local region or evolutionary peculiarities of magnetic stars. A more detailed analysis, involving magnetic field modelling that incorporates rotation periods and comparisons with 3D-models of the Galaxy, will be the subject of future investigations.

\section*{FUNDING}

The author expresses gratitude to I.~A.~Yakunin and A.~V.~Moiseeva for their assistance in the work. The study was carried out with financial support from the Russian Science Foundation (grant \textnumero~25-12-00003).

\label{lastpage}

\end{document}